\documentclass[preprint]{aastex}
\singlespace
\slugcomment{accepted for the publication in the {\it Astrophysical Journal}}
\def\ros{{\it ROSAT}}
\def\asca{{\it ASCA}}

\def\la{\mathrel{\hbox{\rlap{\hbox{\lower4pt\hbox{$\sim$}}}\hbox{$<$}}}}
\def\ga{\mathrel{\hbox{\rlap{\hbox{\lower4pt\hbox{$\sim$}}}\hbox{$>$}}}}

\shortauthors{Park}
\shorttitle{Galactic X-ray Background}

\begin{document}
\title{ASCA Observation of an ``X-ray Shadow'' in the Galactic Plane}
\author{\bf Sangwook Park and Ken Ebisawa}
\affil{Laboratory for High Energy Astrophysics, Code 662 \\
NASA/Goddard Space Flight Center, Greenbelt, MD. 20771 \\
and \\
Universities Space Research Association} 

\email {spark@lobster.gsfc.nasa.gov,ebisawa@olegacy.gsfc.nasa.gov}

\begin{abstract}
The diffuse X-ray background (DXB) emission near the Galactic plane
($l,b~\sim~25.6^{\circ},0.78^{\circ}$) has been observed with $ASCA$. The 
observed region is toward a Galactic molecular cloud which was recently 
reported to cast a deep X-ray shadow in the 0.5 $-$ 2.0 keV band
DXB. The selection of this particular region is
intended to provide a constraint on the spatial distribution
of the DXB emission along the line of sight: i.e., the molecular cloud
is optically thick at $<$2 keV and so the bulk of the observed soft X-rays 
{\it must} originate in the foreground of the cloud, which is at $\sim$3
kpc from the Sun. In the 0.8 $-$ 9.0 keV band, the observed spectrum is 
primarily from multiple components of thermal plasmas. We here report 
a detection of soft X-ray (0.5 $-$ 2 keV) emission from an 
$\sim$10$^{7}$ K thermal plasma. Comparisons with the {\it ROSAT} data 
suggest that this soft X-ray emission is absorbed by $N_H$ = 1 $-$ 3 
$\times$ 10$^{21}$ cm$^{-2}$, which implies a path-length through the 
soft X-ray emitting regions of $\la$1 kpc from the Sun.
\end{abstract}


\section {\label {sec:intro} INTRODUCTION}

The 0.1 $-$ 0.3 keV band diffuse X-ray 
background (DXB) emission in the Galactic plane has been attributed to 
the emission from the Local Hot Bubble (LHB):
an $\sim$10$^{6}$ K plasma filling an extensive cavity, 
where absorbing neutral material is difficient, with an average
radius of $\sim$100 pc around the solar system \citep{cr87,sel98}. 
At $>$2 keV it has been known that there
exists unresolved DXB emission along a thin disk of the plane 
($-$60$^{\circ}$ $<$ {\it l} $<$ 60$^{\circ}$): the
so-called Galactic ridge X-ray emission (GRXE) \citep{wel82,wel85}.
The detection of the Fe K line emission, with a scale height of
$\sim$100 pc \citep{yk93}, revealed the thermal origin of
the GRXE \citep{kel86a}. Unresolved point sources would be 
insufficient to produce the observed X-ray flux \citep{yel97,ysel97},
and the bulk of the GRXE appears to have diffuse origins such as 
multiple supernova remnants \citep{kel86b,kel97,vm98}.
The implied temperature for the thermal plasma to make up the 2 $-$ 10
keV GRXE ranges $\sim$10$^{7.5}$ $-$ $\sim$10$^{8}$ K \citep{kel86a,kel97,
yel97,vm98}.

The nature of the 0.5 $-$ 2 keV band DXB along the plane is elusive.
The presence of a few million K Galactic gas beyond the LHB and nearby
supernova remnants has been suggested in order to incorporate the observed
soft X-ray emission in the plane \citep{nel82}. The distribution
and the spectral properties of the X-ray emitting material along the line 
of sight have yet to be known.
One of the major advances in the study of the 0.5 $-$ 2 keV band
Galactic DXB in the plane came with the detection of
``X-ray shadows'' cast by distant molecular clouds. 
The X-ray shadows in the DXB as detected
with mosaics of the {\it ROSAT} PSPC pointed observations in the Galactic
plane have recently revealed the existence of a highly enhanced X-ray
emitting region (i.e., the derived X-ray background intensity beyond the 
absorbing molecular clouds is more than an order of magnitude brighter 
than the nominal high latitude intensity) around the Galactic center:
the Galactic X-ray bulge (GXB) \citep[P98 hereafter]{pel97,ael00,pel98}.
The estimated plasma temperature of the GXB is 
$\sim$10$^{6.6 - 6.7}$ K. These results are in good agreement with the
\ros~all-sky survey (RASS) data analysis in the general direction of
the Galactic center~\citep{sel97}. P98 further suggested that the angular
extension of the GXB along the first quadrant of the plane may be 
confined by the molecular ring at {\it l} $\sim$ 25$^{\circ}$. 

The results of these X-ray shadow studies also
suggest that there exists a substantial fraction of DXB emission
originating foreground to the absorbing molecular clouds, which are
located at 2 $-$ 3 kpc from the Sun. The ``foreground component''
was tentatively attributed to the emission from a $\sim$10$^{7}$ K plasma
based on a band-fraction analysis \citep{p98}. 
The extensive spectral analyses of
individual DXB components along the line of sight were however 
infeasible with the {\it ROSAT} PSPC data due to the limited spectral 
capabilities of the PSPC mosaics.

In order for a spectral study of the Galactic
DXB originating foreground to the X-ray shadow clouds,
we carried out an {\it ASCA} observation in the 
direction of {\it l, b} = 25.6$^{\circ}$, 0.78$^{\circ}$. 
The {\it l, b} = 25.6$^{\circ}$, 0.78$^{\circ}$
direction is toward a Galactic molecular cloud which was recently
reported to cast an X-ray shadow in the 0.5 $-$ 2.0 keV band (P98).
The existence of the dense molecular cloud with a well-established kinematic
distance (d $\sim$ 3 kpc) is expected to provide a useful constraint
for the spatial distribution of the DXB emission along the line of sight. 
In other words, in the soft energy band ($<$2 keV), the molecular
cloud is optically thick and the distance to the cloud ($\sim$3 kpc) sets
a {\it limit} for the X-ray emitting region, or  regions, along the 
line of sight.
The contribution from unresolved stellar sources to the observed 0.5 $-$ 2 keV
band DXB in the plane has been shown to be small \citep{ss90,w92,os92}.
{\it ASCA} observations of the Scutum arm region ({\it l} $\sim$
28$^{\circ}$ $-$ 29$^{\circ}$) have also demonstrated that only $\sim$1\%
of the DXB emission in the plane is from unresolved point sources
\citep{yel97}. The bulk of the DXB in the plane is therefore
{\it truly} diffuse. We can thus perform 
the spectral study of the 0.5 $-$ 2 keV Galactic DXB emission, the bulk 
of which is diffuse and originates within $\sim$3 kpc from the Sun.
Utilizing the full effective
energy range of \asca, the overall spectrum is analyzed and is 
compared with the previous studies of the GRXE
which were performed at different directions in the sky.
The observation is described in \S\ref{sec:obs}.  
The analysis and results are presented in \S\ref{sec:analysis} and the
implications are discussed in \S\ref{sec:discussion}. A summary and the
conclusions are presented in \S\ref{sec:summary}.

\section{\label{sec:obs} OBSERVATION}

\begin{figure}[]
\figurenum{1}
\plottwo{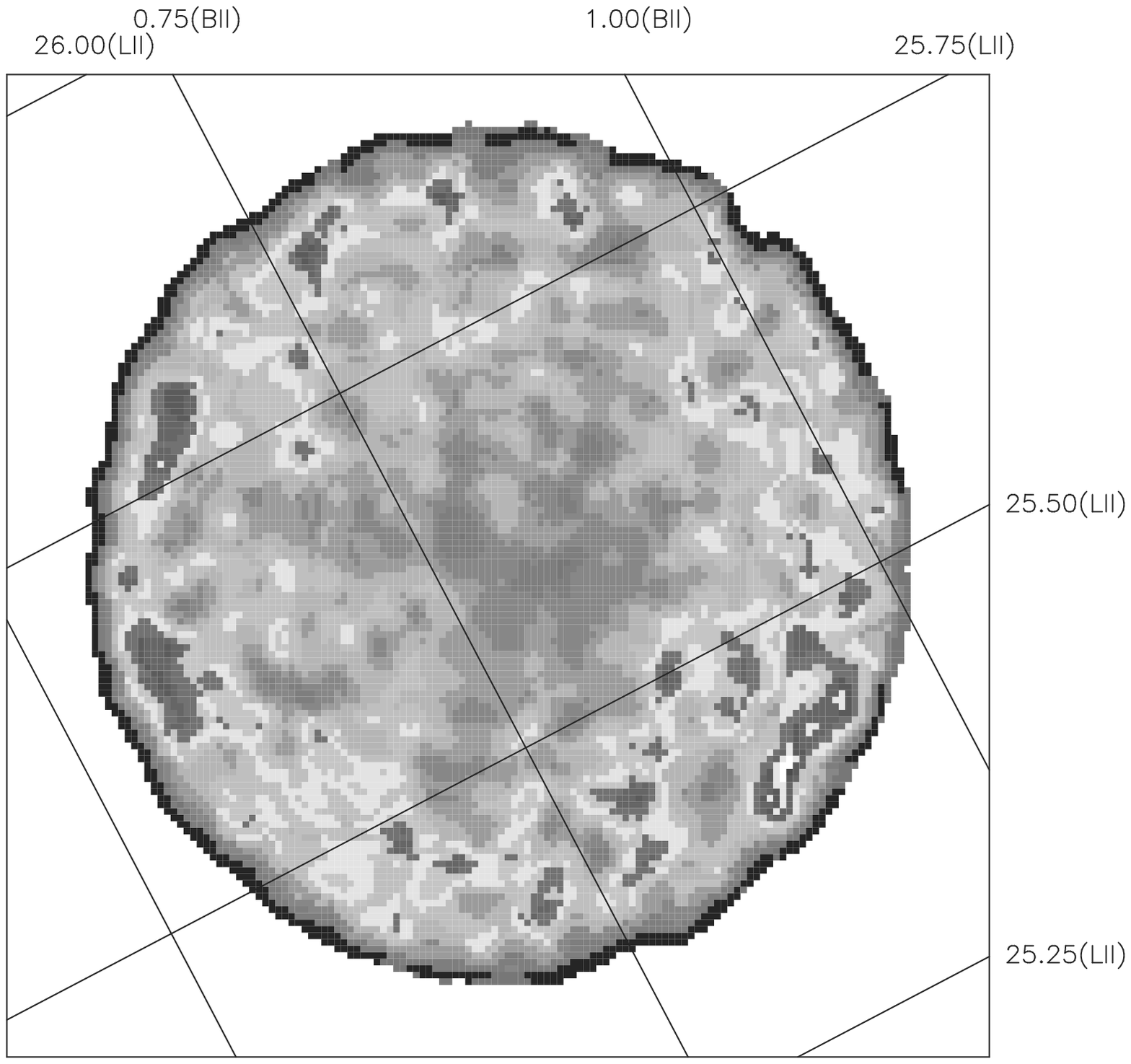}{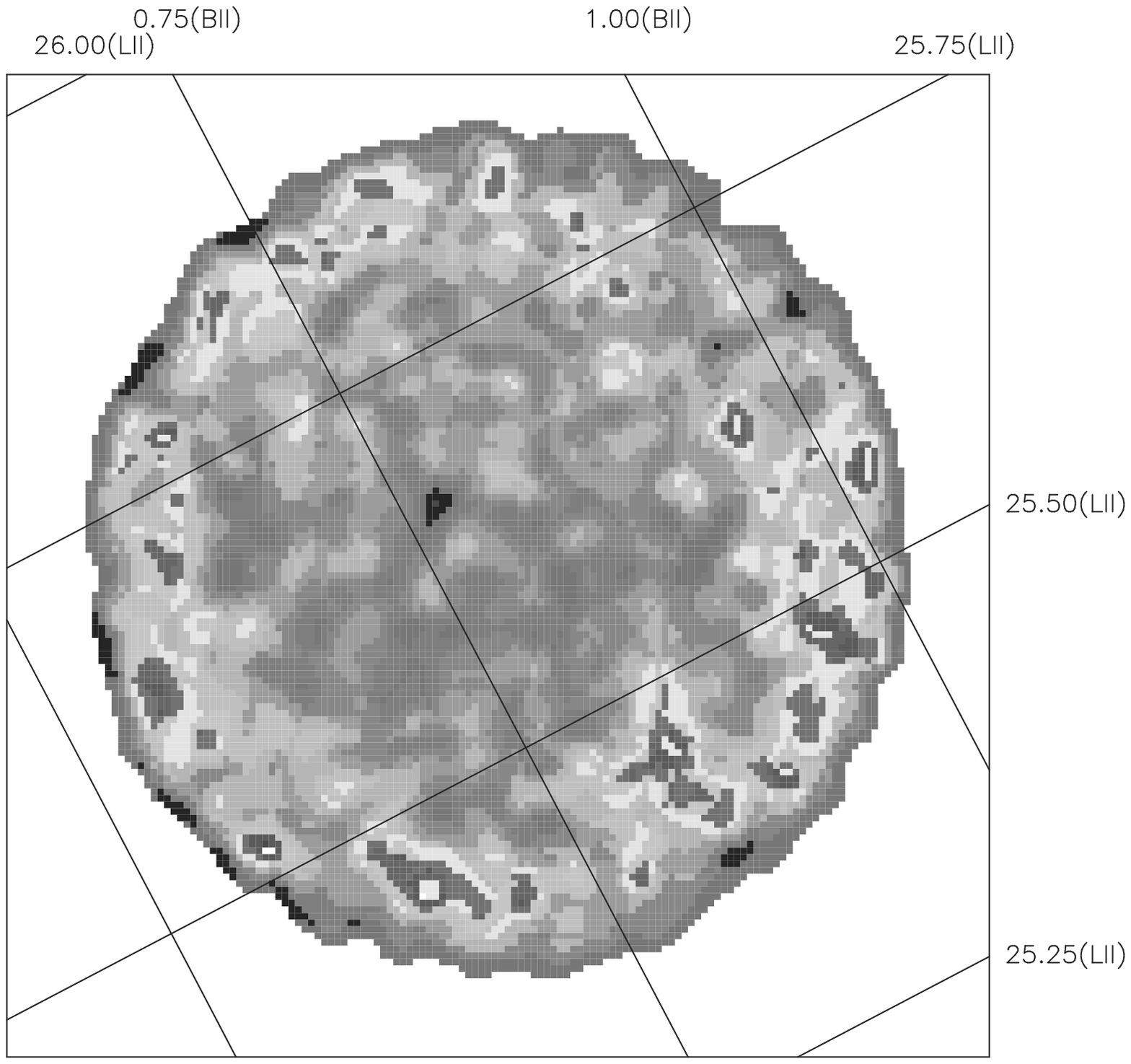}
\figcaption[]{
{\it Left}: (a) The exposure and vignetting corrected 2.0 -- 9.0 keV band 
{\it ASCA} (GIS2+GIS3) image. 
{\it Right}: (b) The exposure and vignetting corrected 0.8 -- 2.0 keV band 
{\it ASCA} (GIS2+GIS3)
image. Both images are smoothed with a Gaussian with $\sigma$=2 pixels.
\label{fig:fig1}}
\end{figure}

The {\it ASCA} observation in the direction of $l,b~=~25.6^{\circ},
0.78^{\circ}$ was carried out on 1998 March 31 during the AO6 phase.
The data from the Gas Imaging Spectrometer (GIS) were obtained in the
pulse-height (PH) mode and the total exposure for each detector
(GIS2 and GIS3) is $\sim$38 ks.
For the analysis the data are integrated over two detectors
for the inner 30$'$ diameter circular region. 
The pointing direction ($l,b~=~25.6^{\circ}, 0.78^{\circ}$) is centered
on a Galactic molecular cloud which has an angular extent of 
$\sim$1$^{\circ}$ to cast a deep X-ray shadow 
in the 0.5 $-$ 2.0 keV band (P98).
The exposure and vignetting corrected GIS images are displayed in
Figure~\ref{fig:fig1}. Figure~\ref{fig:fig1}a is the hard band (2.0 $-$
9.0 keV) image and Figure~\ref{fig:fig1}b is the soft band (0.8 $-$ 2.0
keV) image of the inner 30$'$ diameter region. 
The observed count rate is 0.18 $\pm$ 0.002 counts s$^{-1}$ 
within 30$'$ diameter in the 0.8 $-$ 9.0 keV band. 
There is a faint point-like source detection within 30$'$ 
diameter, at RA(J2000)=18$^{h}$ 34$^{m}$ 52$^{s}$.3, 
Dec(J2000)=$-$06$^{\circ}$ 16$'$ 37$"$.1,
in the 0.8 $-$ 9 keV band. The source counts within 3$'$ radius
detection circle is, however, small ($\sim$1\% of the total counts).

The Solid-state Imaging Spectrometer (SIS) imposes some problems for the
study of faint diffuse background emission. 
The SIS has experienced a decrease in detector efficiency and an 
increasing divergence from the GIS spectra since the early phase of the 
mission due to the accumulated radiation damage and the increasing 
fluctuation of the dark current. This degradation of the SIS performance
can be substantial for the faint X-ray background spectra 
since the position depenedence of the SIS performance change is not 
known for the {\it diffuse} background emission. 
The SIS also has a smaller effective area 
which also makes it inappropriate for the study of the faint 
diffuse background emission.
The count rate for the SIS is $\sim$0.07 counts s$^{-1}$ even with the
full effective solid angle ($\sim$480 arcmin$^{2}$), which is only 
$\sim$39\% of the GIS count rate of the inner 30$'$ diameter region. 
Considering that the photons detected near the edge of the nominal field
of view may not be useful for the analysis, the number
of utilizable photons with the SIS would be smaller than that of the GIS 
(30$'$ diameter) at least by a factor of 4 $-$ 5.
In this paper we thus present only the GIS result which is free from 
systematic uncertainties and more reliable than SIS.

\section{\label{sec:analysis} ANALYSIS \& RESULTS}

\subsection{\label{subsec:image} X-ray Images}

The observed X-ray emission is smooth with no significant contribution 
from point sources (Figure~\ref{fig:fig1}). In order to investigate the 
distribution of the X-ray intensity over the field of view, radial 
profiles of the observed count rates for the non-cosmic X-ray background 
(NXB), data, and the cosmic X-ray background (CXB) components are compared. 
The NXB counts can be extracted with the night-Earth data. The night-Earth 
data are available from the HEASARC {\it ASCA} Guest Observer Facility 
(GOF), which is an accumulation of the data from 1993 June to 1999 August. 
Since the night-Earth background is known to have long-term variations 
over the years, the data taken only between 1998 January and 1998 May are 
extracted in order to accomodate the observation dates of our data. 
The CXB can be obtained with the ``blank-sky'' data released by the 
{\it ASCA} GOF at HEASARC, which are point-source removed high latitude 
observations taken between 1993 June and 1995 December. Since the CXB 
is presumably isotropic over the entire sky with no intensity variations, 
the blank-sky data can be utilized to test the angular distribution of 
our data. The observed data, night-Earth and blank-sky data are 
extracted in the soft (0.8 $-$ 2 keV), the hard (2 $-$ 9 keV), and
the total (0.8 $-$ 9 keV) band and the count rates are integrated over 
the annular bins of 1.5$'$ within the 15$'$ radius circular region. 

\begin{figure}[]
\figurenum{2}
\epsscale{0.7}
\plotone{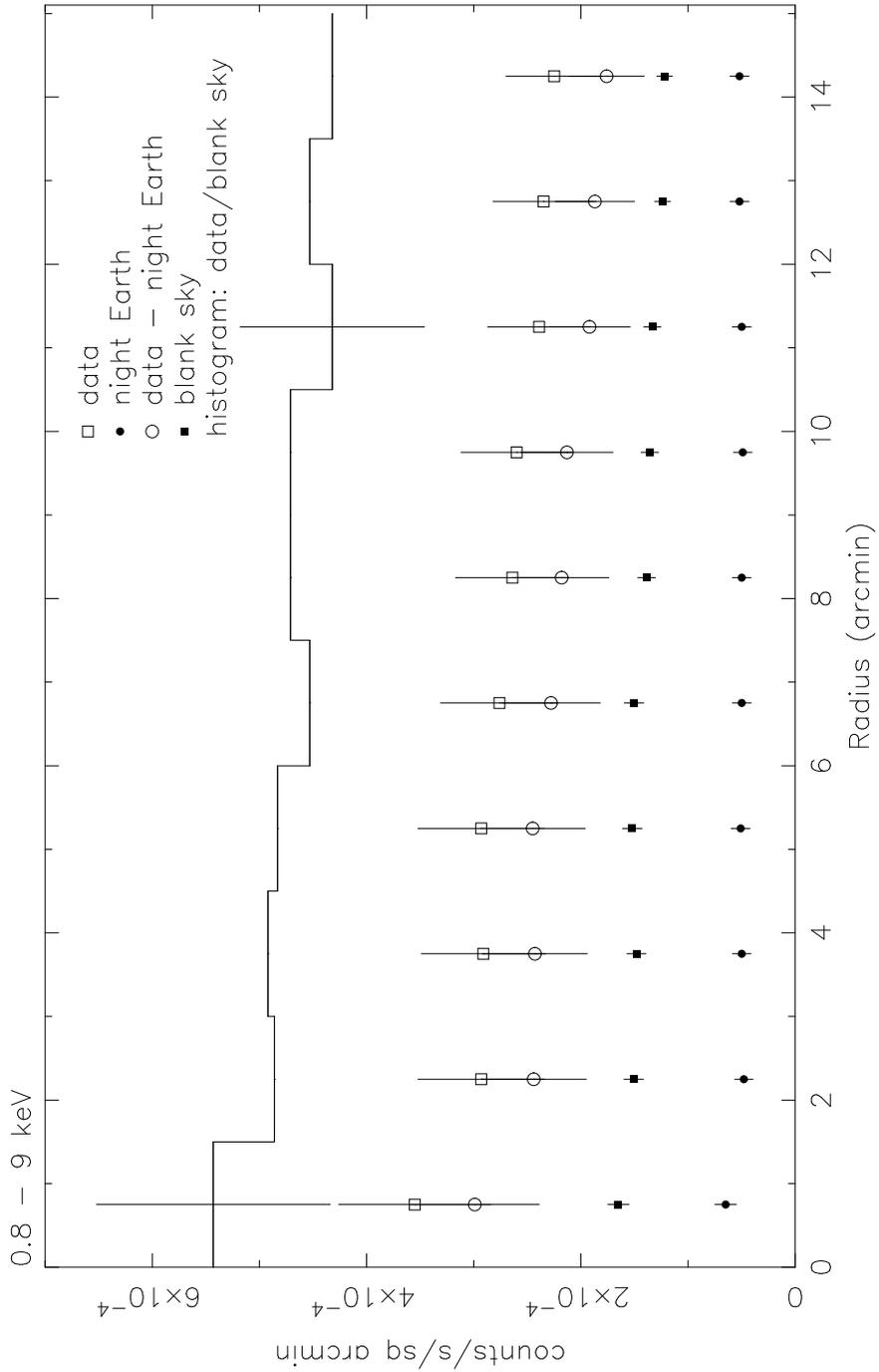}
\figcaption[]{
Radial profiles of the 0.8 $-$ 9 keV DXB from the GIS. The count rates
are integrated over the annular bins of 1.5$'$ within the 15$'$
radius region. The night-Earth subtracted data to blank-sky ratio as
presented with histograms has been arbitrarily scaled to fit the display
window. The vertical lines on the histogram present typical errors.
\label{fig:fig2}}
\end{figure}

The radial profiles of the 0.8 $-$ 9 keV band average
count rates in each bin are displayed in Figure~\ref{fig:fig2}. 
In Figure~\ref{fig:fig2}, the open-squares are observed data, 
filled-circles are the night-Earth count rates, open-circles are (data $-$ 
night-Earth), and filled-squares are blank-sky. Although the night-Earth 
background generally has higher intensity around the edge of the detector
\citep{ish96}, the effect is negligible within 15$'$ from the center
(Figure~\ref{fig:fig2}). 
Assuming the CXB intensity distribution is uniform over the sky, the
intensity distribution of the X-ray data can be investigated
by calculating the intensity ratio between the observed data and the 
blank-sky data. In Figure~\ref{fig:fig2}, the histograms present the
radial profiles (arbitrarily scaled) of the data to blank-sky intensity 
ratio where the night-Earth contribution is subtracted from both. Within 
15$'$ radius of the nominal field of view, no statistcally significant 
intensity variation is observed. The same analysis has been performed
in the soft and the hard band. The results from these subbands
also indicate no significant intensity variations over the field.

Since the effect of the shadow is not explicitly seen in the images,
off-cloud observations may be necessary to assure the utility of the
X-ray shadow in constraining the distribution of the soft DXB emission
along the line of sight. No such observations are currently available
leaving it for follow-up observations. The solid angle of our data is
however only 30$'$ diameter and is {\it completely} within the angular 
extent of the shadow. No intensity variation in the soft band image is 
thus not surprising considering no significant column variations around 
the central region of the cloud (P98) and the limited field of view 
for the data covering only the central region of the shadow. The smooth 
distribution in the hard band image may also be reasonable since
the cloud is most likely {\it transparent} on average in the 
2 $-$ 9 keV band ($\tau$ $\ll$ 1). There may be some contribution 
from the off-cloud emission due to the ``stray-light'' effect 
(see \S \ref{subsec:arf}). For example, the contribution from offset 
angle $>$ 1$^{\circ}$ is expected to be $\la$ 15\% of the total counts 
in the 1 $-$ 2 keV band (the contribution in the hard band is expected
to be smaller) \citep{ish96}. The bulk of the observed flux 
in the soft band is thus assumed to arise in on-cloud region presumably 
representing the X-ray emission originating in the foreground of the cloud.

\subsection{\label{subsec:arf} Background subtraction \& Energy response }

The study of the diffuse background emission with \asca~is substantially
affected by the {\it stray light}, which is a ``leakage'' of the photons
from outside of the nominal field of view due to the broadly extended
{\it ASCA} mirror responses \citep{sel95}. A specific energy response of 
the detector must be generated to resolve this problem. For this purpose 
an ancillary response file (ARF) for the diffuse emission is generated 
through ray-tracing, assuming an {\it uniform} X-ray emitter having 
1.5$^{\circ}$ radius. This technique is essentially identical
to those by Kaneda et al. (1997) and Valinia et al. (2000). Although 
the diffuse ARF is created for the Galactic diffuse emission, it is also 
used in determining the normalization for the extragalactic power law. 
This specific ARF is utilized in the following spectral analysis.

As the first step of the spectral analysis the NXB is subtracted using 
night-Earth data. The extragalactic power law 
component of the DXB (presumably from unresolved AGNs) then needs to be
considered since the contribution from this component in the photon 
energy of $>$2 keV is significant even in the plane.
The power law component of the DXB has been extensively studied with 
high latitude data and can be described with a photon index of
$\Gamma$ $\sim$ 1.4 in the 2 $-$ 10 keV band \citep{gen95,cfg97}.
This extragalactic power law component is included in the spectral fitting
as a fixed component. In order to produce an appropriate normalization of 
the power law component, blank-sky data are fitted with a
simple power law model in the 0.8 $-$ 9 keV band after subtracting
the NXB. The best fit photon index
is $\Gamma$ = 1.46 (${\chi}^{2}_{\nu}$ = 1.69) which is consistent with
previous results \citep{ish96}. 
A range of absorbing column density N$_{H}$ = 3 $-$ 5 
$\times$ 10$^{22}$ cm$^{-2}$ should be considered for the power law component 
accounting for the absorption by the Galactic plane asumming an average 
space density of 1 cm$^{-3}$ in the Galactic disk. 
We found that the difference due to the assumed N$_{H}$ values
in this range does not affect the fit, and a fixed value of 
N$_{H}$ = 4 $\times$ 10$^{22}$ cm$^{-2}$ is used for the analysis.

\subsection{\label{subsec:spec} Spectral Analysis}

\subsubsection{\label{rsmodel} Raymond-Smith Model Fit}

The observed spectrum should be  associated with the GRXE,
considering the observed direction and the effective energy band.
For the study of the GRXE,
optically thin thermal plasmas both in ionization equilibrium and 
non-equilibrium have been considered by previous authors 
\citep{kel86a,ysel97,kel97,yk95,vm98}.
With the {\it Ginga} data
Yamauchi \& Koyama (1995) have shown that the GRXE continuum
temperature of 5 $-$ 10 keV is too high for the detected 6.7 keV Fe
line assuming an ionization equilibrium, 
and suggested a non-equilibrium state for the GRXE.
A non-equilibrium ionozation
model was used by Kaneda et al. (1997) with {\it ASCA} data and the
best fit temperature of $\sim$7 keV was obtained for the hard component.
On the other hand,  presence of a power-law 
hard-tail in the GRXE spectrum in addition to the thermal components
has been suggested \citep{ysel97,vm98}.
As a matter of fact, the GRXE as observed with 
{\it RXTE} was suggested to arise from
an $\sim$2 $-$ 3 keV plasma in an ionization equilibrium as well as
from a power law component of $\Gamma$ $\sim$ 1.8 which dominates at
$>$10 keV \citep{vm98}.
The ionization state of the GRXE has thus been uncertain and the
previous results have been dependent on different assumptions in the
models. 

\begin{figure}[]
\figurenum{3}
\epsscale{0.7}
\plotone{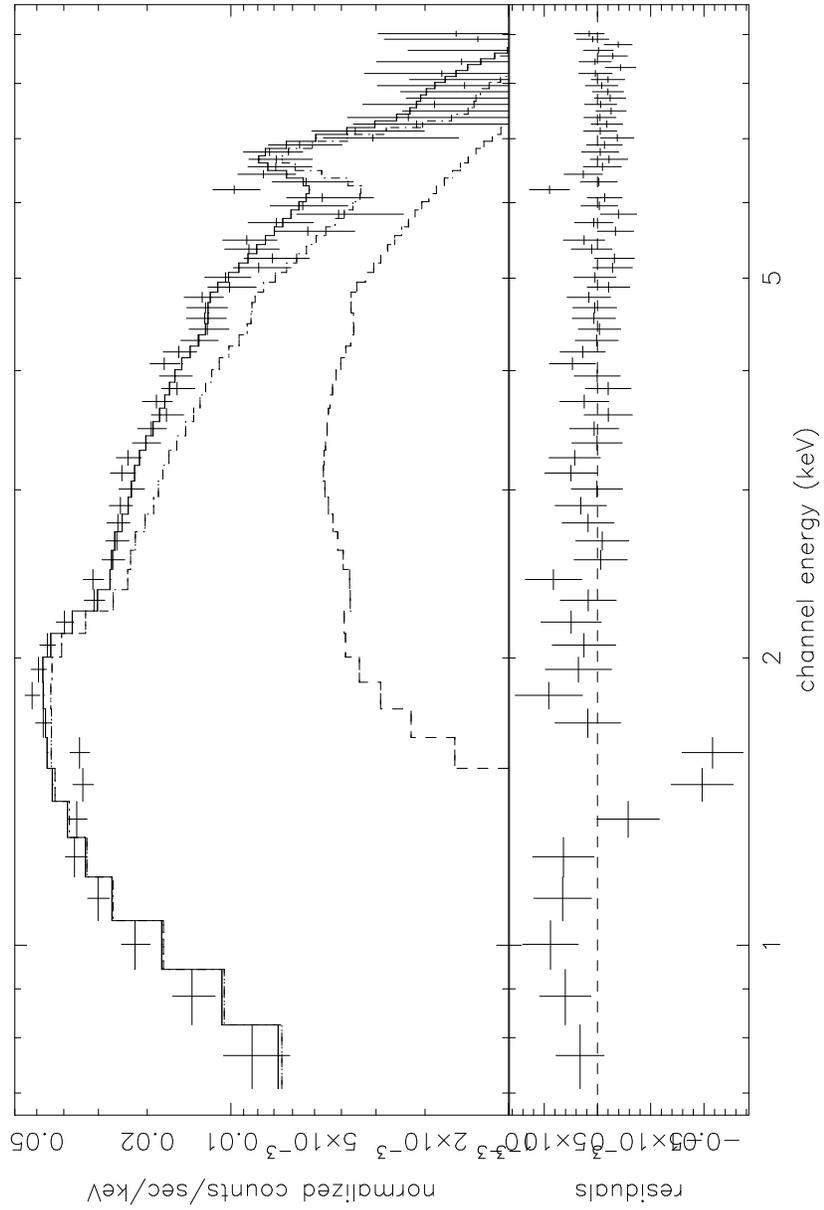}
\figcaption[]{
The ASCA/GIS spectrum folded through the instrumental response 
with the best-fit one component Raymond-Smith model. The dashed 
line represents the fixed extragalactic power law component.
\label{fig:fig3}}
\end{figure}

Considering the complex nature of the modeling GRXE plasma,
in this paper,
we assume that thermal plasmas are in ionization equilibrium,
as a first step toward the understanding of the nature
of ``X-ray shadow''.  Our results in this paper may be subject to the
assumption of  ionization equilibrium, on which further 
investigations will be required.
The spectral fitting is first performed with an absorbed Raymond-Smith (RS)
model assuming thermal equilibrium. The best fit plasma 
temperature is {\it kT} = 6.30$^{+2.58}_{-1.25}$ 
keV with an absorption of $N_H~=~0.55^{+0.13}_{-0.14}~{\times}~10^{22}$ 
cm$^{-2}$ (the errors refer to a 90\% confidence level, hereafter),
where we used standard single absorption model (see \S \ref{cont_abs}).
The best fit model is displayed in Figure~\ref{fig:fig3}.
In Figure~\ref{fig:fig3} the fixed extragalactic power law component is
represented with a dashed line and the contribution is shown to be 
limited at $E~>~2$ keV due to the absorption by the plane.
It is clearly observed that the data are in good agreement with the model
only at $>$2 keV (Figure~\ref{fig:fig3}), and the fit is unacceptable 
(${\chi}^{2}_{\nu}$ = 3.63) at $<$2 keV, the DXB in 
which energy band is motivation of the present work.

A second RS component is then added in the model. 
The fit is now good even at $<$2 keV (${\chi}^{2}_{\nu}$ = 0.92) and the 
overall statistics also improved (${\chi}^{2}_{\nu}$ = 0.59) (Figure
\ref{fig:fig4}). The soft component ({\it kT} = 0.64$^{+0.13}_{-0.16}$ keV
with $N_H$ = 1.21$^{+0.23}_{-0.29}$ $\times$ 10$^{22}$ cm$^{-2}$) 
makes up most ($\sim$75\%) of the 0.5 $-$ 2.0 keV band emission and the 
contribution from this component to the 2 $-$ 10 keV band count rate is 
small ($\sim$6\%). 
The hard component ({\it kT} = 3.71$^{+1.50}_{-0.79}$ keV) dominates the 
observed spectrum at $>$2 keV and produces $\sim$68\% of the count rates 
in the 2 $-$ 10 keV band. The implied absorbing column of the hard component, 
$N_H$ = 
2.47$^{+1.13}_{-0.94}$ $\times$ 10$^{22}$ cm$^{-2}$, indicates that this 
component originates beyond the molecular cloud. 

Even though the overall fit is good, the two component RS model imposes some
descrepancies in the soft band. 
Since the molecular cloud is optically thick ($\tau~\sim~$ 1.7 on average 
between 0.5 and 2 keV, P98) and so casts a deep X-ray
shadow in the 0.5 $-$ 2.0 keV band DXB, the bulk of the soft component
{\it must} originate in the foreground of the cloud. The best fit absorbing
column of the soft component ($N_H~=~1.21~{\times}~10^{22}$ cm$^{-2}$),
however, indicates a distance of $\ga$3 kpc through the emission region
assuming an average space density of 1 cm$^{-3}$ in the plane.
This distance suggests that the soft X-ray emitting region substantially
extends beyond the cloud, which is inconsistent with the
foreground origin for the soft component.

Furthermore, we found that this soft component does not agree with the
{\it ROSAT}\/ data on the same region.
For a comparison with the {\it ROSAT} data, the best fit model is
convolved with the {\it ROSAT} PSPC response and a hardness
ratio is calculated.
The 1.5 keV band (0.73 $-$ 2.04 keV, corresponding to the R6 + R7 band
for the \ros~PSPC) to 0.75 keV band (0.44 $-$ 1.21 keV, corresponding to
the R4 + R5 band for the \ros~PSPC) \citep{sel94} hardness ratio
of the count rates for the modeled spectrum is 2.4. This hardness ratio 
is substantially higher than the observed value, 1.4, of the foreground
count rates of the X-ray shadow in the same region as detected with
\ros~(P98). 

\subsubsection{Continuous Absorption Model}\label{cont_abs}

We have assumed a single ``foreground'' absorbing column for each RS 
component while, in reality, the X-ray emitting plasma and the 
absorbing material is more likely interspersed along the line of sight 
\citep{jk86}. In order to investigate effects of the intermixture of 
the absorbing material and the X-ray emitting plasma, Kaneda (1996) 
has considered the ``multi-absorption model'' and found that the effect in 
the overall {\it ASCA} bandpass was not significant compared to the 
single foreground absorption. 
The ``multi-absorption'' model is still a more realistic picture especially 
in the soft band ($<$ 2 keV) where the absorption is more sensitive to the 
distribution of the absorbing material along the line of sight. 

In the following, 
we thus investigate both of the single and the ``continuous'' absorption 
model for the soft band emission. The single absorption 
is the case that the absorber is in front of the emitter, 
where we take the standard 
interstellar absorption model, such that $\exp(-N_H \sigma_H)$ is multiplied
to the thermal plasma model. Here, $N_H$ is the hydrogen column
density, and $\sigma_H$ is the average cross section per hydrogen
atom by  Morrison \& McCammon (1983). We also consider the case 
that the emitter and absorber are completely intermingled and 
distributed uniformly along the line of sight. 
For this ``continuous absorption model'' we use the formula,
$ (1- \exp(-N_H \sigma_H))/(N_H \sigma_H)$ instead.
It is likely that the true distribution of the emitter and absorber 
will be  somewhere in-between, so the model parameter ranges (such 
as $N_H$) we obtain using the two absorption models are considered 
to be conservative limits. For the hard component 
($>$ 2 keV), where the effect of the distribution of the absorbing material
is not significant, only the single absorption utilizing the standard
interstellar absorption is used in the following analysis.

\subsubsection{Two Bremsstrahlung Models plus Gaussian Line Fits}

\begin{figure}[]
\figurenum{4}
\epsscale{0.7}
\plotone{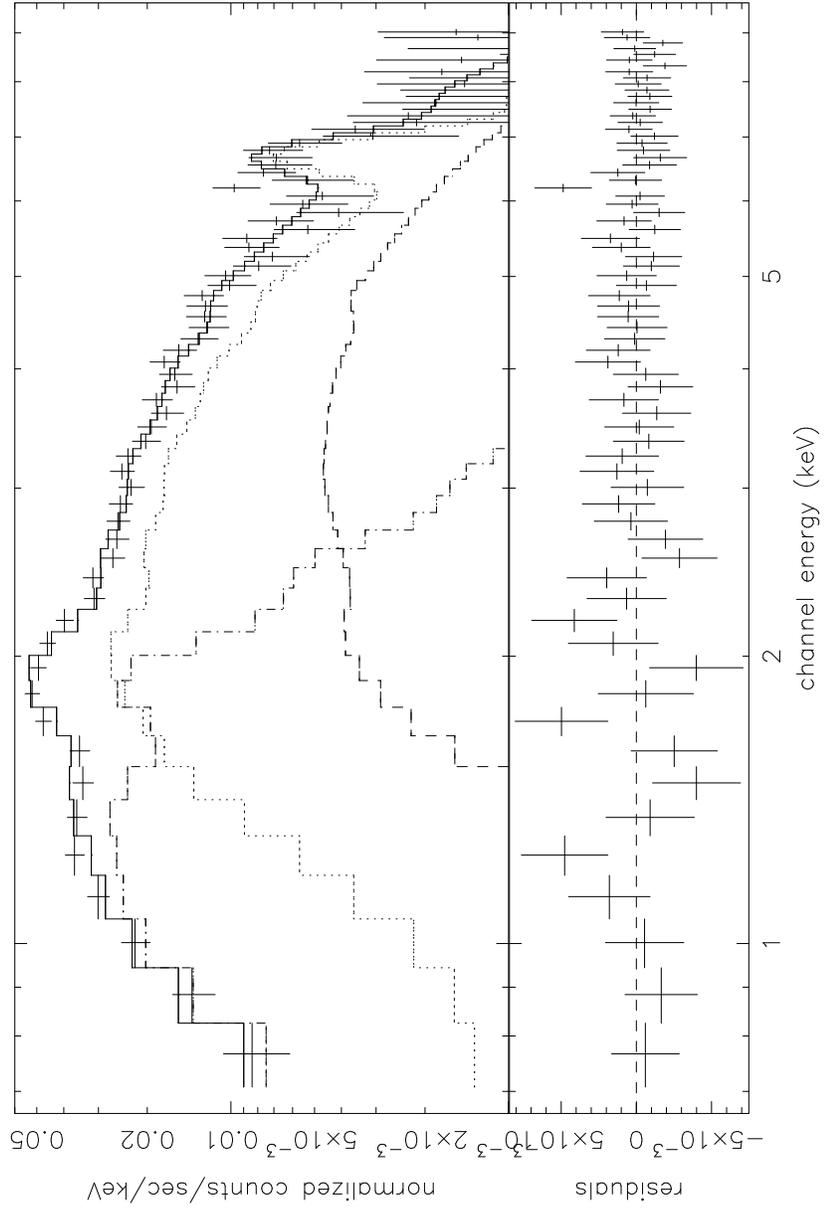}
\figcaption[]{
The ASCA/GIS spectrum folded through the instrumental response 
with the best-fit two component Raymond-Smith model.
The dashed line is the fixed 
extragalactic power law. The dot-dashed line represents the best fit
soft component of the Raymond-Smith and the dotted line 
is the best fit hard component of the Raymond-Smith model.
\label{fig:fig4}}
\end{figure}

The observed soft X-ray spectrum  cannot be 
properly represented  with the two component RS model, because the
required soft component column density is too high and the
hardness ratio is inconsistent with the ROSAT result (\S 
\ref{rsmodel}).  Note that 
the required column density would become even higher if we use the
continuous absorption model (\S \ref{subsec:soft}).
Origin of the descrepancies may be in part due to 
our assumption of the full ionization equilibrium of the plasma.
The soft band X-ray emitting plasma in the plane has been suggested to be 
in a non-equilibrium state \citep{kel97} while we have assumed a 
thermal equilibrium. At least the RS model appears not effective 
to fit ``unresolved'' lines in the soft band spectrum 
 as suggested by a systematic deviation in the
residuals from the best fit model at $E~<~2$ keV (Figure~\ref{fig:fig4}).
As an attempt to improve the soft band spectral fitting, 
we fit the spectrum with two component thermal bremsstrahlung model with
separate Gaussian components for the candidate atomic emission lines.


\label{subsec:line}
Kaneda et al. (1997) have detected several atomic line emission
(e.g., Ne K, Fe L, Mg K, Si K, etc.) at $<$ 2 keV in
the Galactic plane DXB spectra, and our data show marginal evidence 
for emission lines at $\sim$1.8 keV and $\sim$1.3 keV in addition
to the Fe line at $\sim$6.7 keV (Figure~\ref{fig:fig4}).
We thus fitted the observed spectrum with two component bremsstrahlung
and three Gaussians. The best fit temperature for the continuum 
to make up the soft band emission is {\it kT} $\sim$ 10 keV with low 
absorption and the hard band emission is from {\it kT} $\sim$ 2 keV 
plasma being absorbed by $N_H$ $\sim$ 3 $\times$ 10$^{22}$ cm$^{-2}$. 
Although the overall fit is good (${\chi}^{2}_{\nu}$ = 0.48), the best 
fit parameters appear unphysical by presenting a very high temperature 
plasma with a small absorption and a lower temperature plasma with a 
large absorption. This is unrealistic, since in the 
neighborhood of solar system, such a high temperature diffuse emission
has not been detected.  Also, this result contradicts with
the more extensive observations of the GRXE by Kaneda et al.\ (1997), such that
low temperature component is dominant in the lower energy band and
so is the high temperature component in the higher energy band.

Although a simultaneous fitting with both soft and hard components
is supposed to be ideal, such fitting is infeasible with low statistics 
of our data as we discussed above. We therefore decompose the observed
spectrum in the hard and soft band to fit separately. Since the hard
component shows a simpler spectral structure and has been more 
extensively studied by previous authors, we fit the hard component
first and then fit the soft component with the hard component
parameters fixed at the best fit values.
For the hard component fitting a conservative cut-off at 3 keV,
where the hard component dominates the observed GRXE spectrum
\citep{kel97}, is taken for our data; i.e.,
in the 3 $-$ 9 keV band, the spectrum is fitted with a thermal 
bremsstrahlung and a Gaussian line in order to determine the center 
energy of the Fe K line.
The best fit line energy is 6.62$^{+0.13}_{-0.14}$ keV where the continuum
temperature is {\it kT} = 3.67$^{+4.16}_{-1.67}$ keV with an absorption of 
$N_H$ $\sim$ 2.1 $\times$ 10$^{22}$ cm$^{-2}$ (${\chi}^{2}_{\nu}$ = 0.45). 
The temperature and the absorption values seem
reasonable since previous observations indicate that the GRXE at 
2 $-$ 10 keV is from a {\it kT} = a few keV plasma 
and that is distributed along a thin disk of the plane \citep{ysel97,kel86a,kel97}. 
The line width was fixed at zero assuming a narrow emission line. When the
line width is allowed to vary the best fit value is 233 eV with the uncertainty
covering zero, which makes the narrow line assumption reasonable.

\begin{figure}[]
\figurenum{5}
\epsscale{0.7}
\plotone{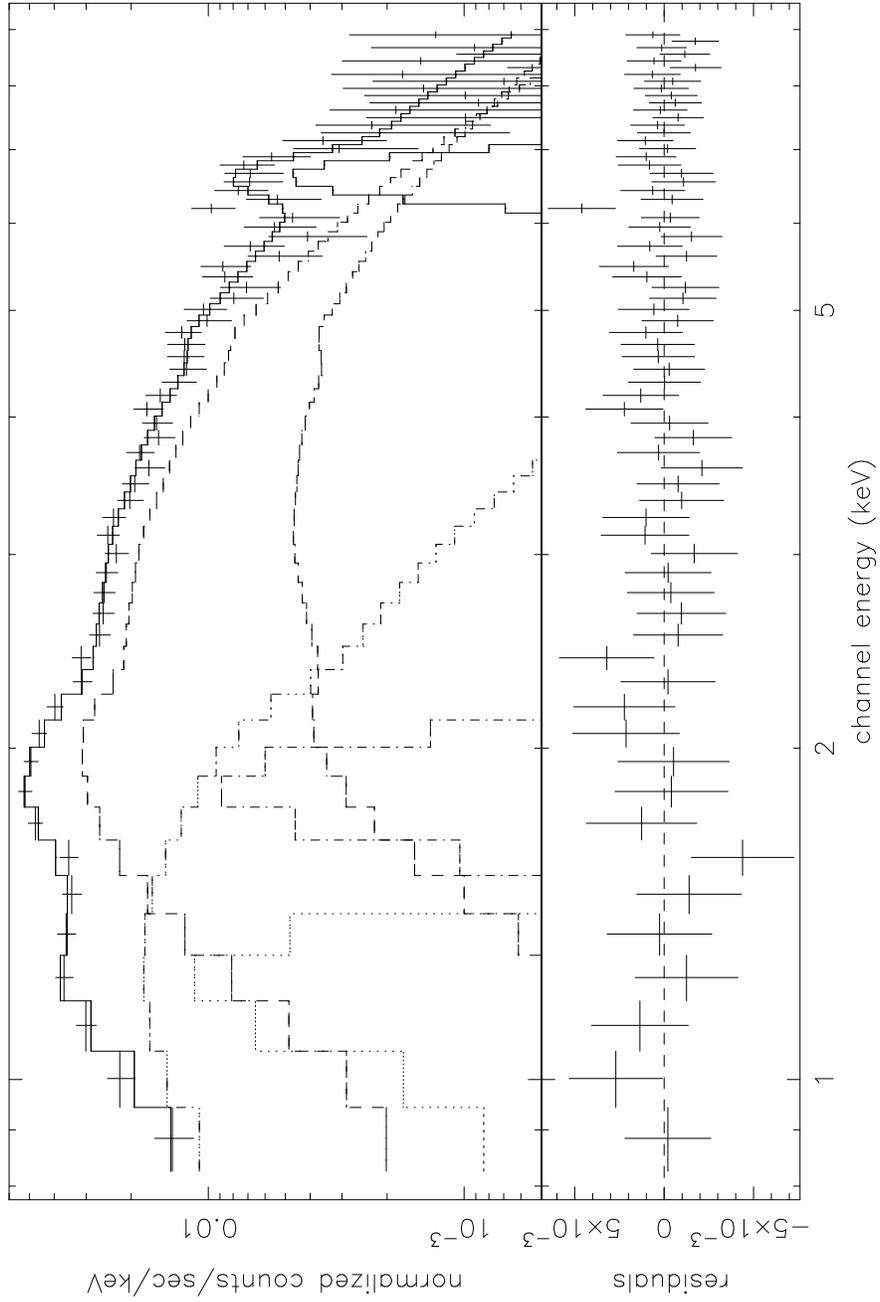}
\figcaption[]{
The ASCA/GIS spectrum folded through the instrumental response 
with the best-fit two component thermal bremsstrahlung (with the
continuous absorption for the soft component). Candidate 
atomic emission lines are fitted with separate Gaussians. 
\label{fig:fig5}}
\end{figure}

After determining the continuum temperature for the hard component and the 
Fe K line energy, the line energies and continuum temperature for the soft 
band emission are fitted.
Another bremsstrahlung model and two Gaussian components 
are added in the 0.8 $-$ 9 keV band. 
%
The best fit line energies are 1.83$^{+0.07}_{-0.06}$ keV and
1.23$^{+0.05}_{-0.08}$ keV. The line widths are again set to zero assuming
narrow lines as well as to help constrain fitted parameters.
The best fit continuum temperature is {\it kT} = 0.92$^{+0.41}_{-0.39}$ keV. 
The associated absorption is negligible and a 2$\sigma$ upper limit of 4.8 
$\times$ 10$^{21}$ cm$^{-2}$ is obtained with the {\em continuous absorption}\/
model. This implies 
a relatively nearby distance scale (probably $<$1.5 kpc, assuming a space 
density of 1 cm$^{-3}$) through the X-ray emitting regions, which is in good 
agreement with the assumption of the foreground (to the cloud) origin of 
the soft component. The overall fit is good with ${\chi}^{2}_{\nu}$ = 0.45. 
The best fit model is displayed in Figure~\ref{fig:fig5} 
and the model paramters are presented in Table~\ref{tbl:tab1}.
In Table~\ref{tbl:tab1} the results from the same model fitting with the
{\em single absorption}\/ for the soft component are also presented. 
The difference in the best fit 
parameters between the two models are insignificant.
The 1.83 keV line can be identified with the Si K line. The 
identification of the ``1.23 keV line'' is unclear and may represent a blend 
of unresolved lines such as Mg K (1.35 keV) and Fe L (1.08 keV) lines as
detected by Kaneda et al. (1997) toward the Scutum arm region.

\subsubsection{Parameters of the Soft and Hard Components}

The soft component (with the continuous absorption) makes up 76\% of 
the total count rate in the 0.5 $-$ 2.0 
keV band. After deabsorbing the spectrum, the 0.5 $-$ 2.0 keV band 
surface brightness is 2.36 $\times$ 10$^{-8}$ ergs s$^{-1}$ cm$^{-2}$
sr$^{-1}$. Assuming 1 kpc of the emitting region along the line of
sight, the volume emissivity is 9.6 $\times$ 10$^{-29}$ ergs 
s$^{-1}$ cm$^{-3}$. The total luminosity is 1.8 $\times$ 10$^{36}$
ergs s$^{-1}$ assuming a disk-like emission region with thickness of 200 pc
and a radius of 1 kpc. 
The modeled plasma parameters with the single absorption for the soft 
component are almost identical to those with the continuous absorption 
and thus are not separately presented in this paper.

The best fit parameters for the hard continuum are {\it kT} = 3.67 keV, 
with $N_H$ = 2.1 $\times$ 10$^{22}$ cm$^{-2}$. 
The hard component is responsible for 70\% of the total count rate in the 
2 $-$ 10 keV band. After removing the absorption, the estimated 
surface brightness is 1.1 $\times$ 10$^{-7}$ ergs s$^{-1}$ cm$^{-2}$ 
sr$^{-1}$ in the 2 $-$ 10 keV band.
Assuming $\sim$15 kpc for the depth of the emitting region along
the line of sight, the volume emissivity can be estimated to be
$\sim$2.99 $\times$ 10$^{-29}$ ergs s$^{-1}$ cm$^{-3}$. For the
emission volume of $\sim$1.8 $\times$ 10$^{66}$ cm$^{3}$ (assuming a 
thin disk with a radius of 10 kpc and a thickness of 200 pc), the 2 $-$ 10 keV 
total luminosity is estimated to be $\sim$5.4 $\times$ 10$^{37}$ ergs s$^{-1}$.

\section{\label{sec:discussion} DISCUSSION}

The RS model appears not successful in describing the data and
we found that a two component bremsstrahlung model with 
separate Gaussians for the candidate emission lines are
more successful to describe our data.
Although the inclusion of simple Gaussians in the model is rather 
arbitrary and may not adequately represent the complicated spectral 
structure, the overall spectrum is the most effectively described with 
the separate Gaussian components. The fitting with separate Gaussians 
thus likely help represent the observed spectrum at least as an approximation. 
In the following sections the results are, therefore, discussed based on
the model with separate Gaussians and two component bremsstrahlung. 

\subsection{\label{subsec:soft} The Soft Component}

The best fit plasma temperature for the soft component is {\it kT} = 
0.92 keV (continuous absorption) $-$ 1.02 keV (single absorption), 
which is higher than that of the GXB ({\it kT} = 0.35 keV)
\citep{sel97}. Park (1998) has suggested that the foreground DXB emission 
(originating at $\la$3 kpc) of the 0.5 $-$ 2 keV band X-ray shadows in the 
{\it l, b} = 10$^{\circ}$, 0$^{\circ}$ and {\it l, b} = 25$^{\circ}$, 
0.5$^{\circ}$ directions may arise from a {\it kT} $\sim$ 0.9 keV plasma. 
Kaneda et al. (1997) demonstrated that the soft component
of the GRXE toward the Scutum arm region ({\it l} = 28$^{\circ}$
$-$ 29$^{\circ}$) has a temperature of {\it kT} = 0.8 keV.
The present results are in good agreement with these previous works.
Due to the opacity of the absorbing molecular cloud, the bulk of
the soft component can be assumed to arise in the foreground of the cloud. 
The modeled absorbing column is small ($<$ 5 $\times$ 10$^{21}$ 
cm$^{-2}$) and is in good agreement with this assumption. 

\begin{figure}[]
\figurenum{6}
\epsscale{0.7}
\plotone{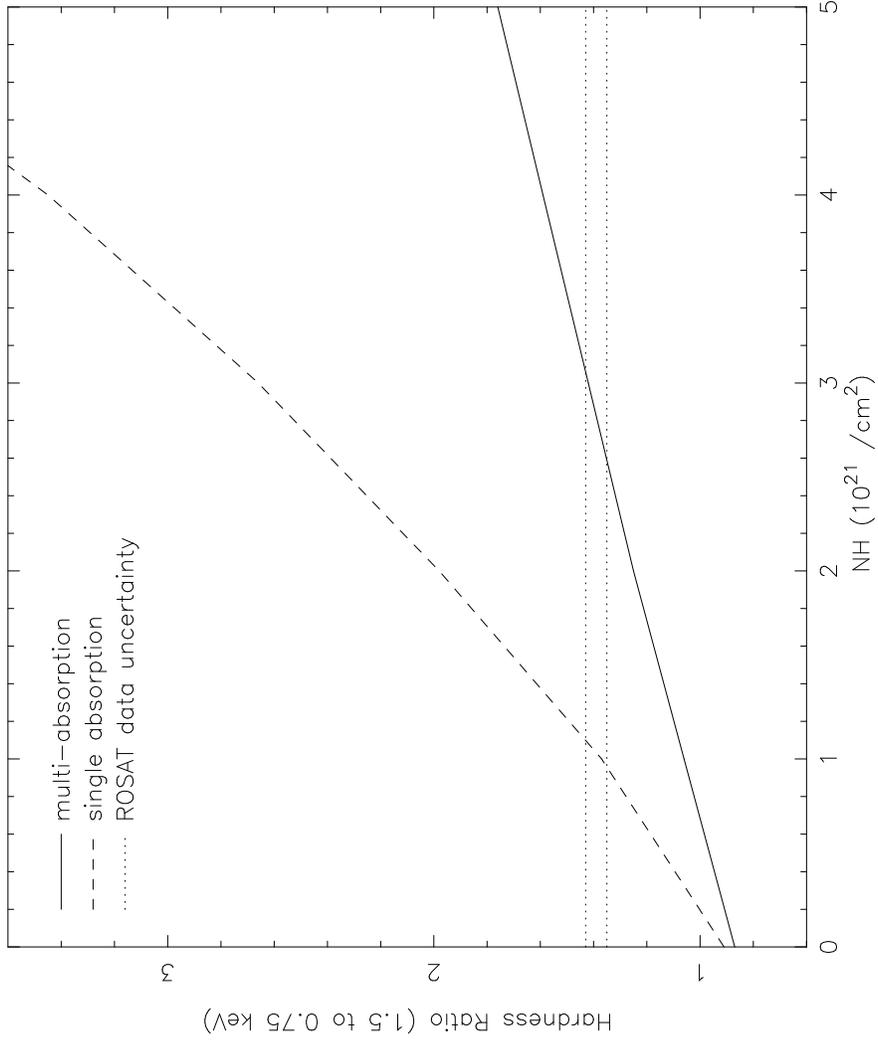}
\figcaption[]{
The modeled 1.5 to 0.75 keV hardness ratios. The solid line is 
for the continuous absorption model, and the dashed line is for
the single absorption model. The area between the horizontal 
dotted lines presents the {\it ROSAT} measurement for the same 
region with embeded uncertainty.
\label{fig:fig6}}
\end{figure}

\begin{figure}[]
\figurenum{7}
\epsscale{1.0}
\plottwo{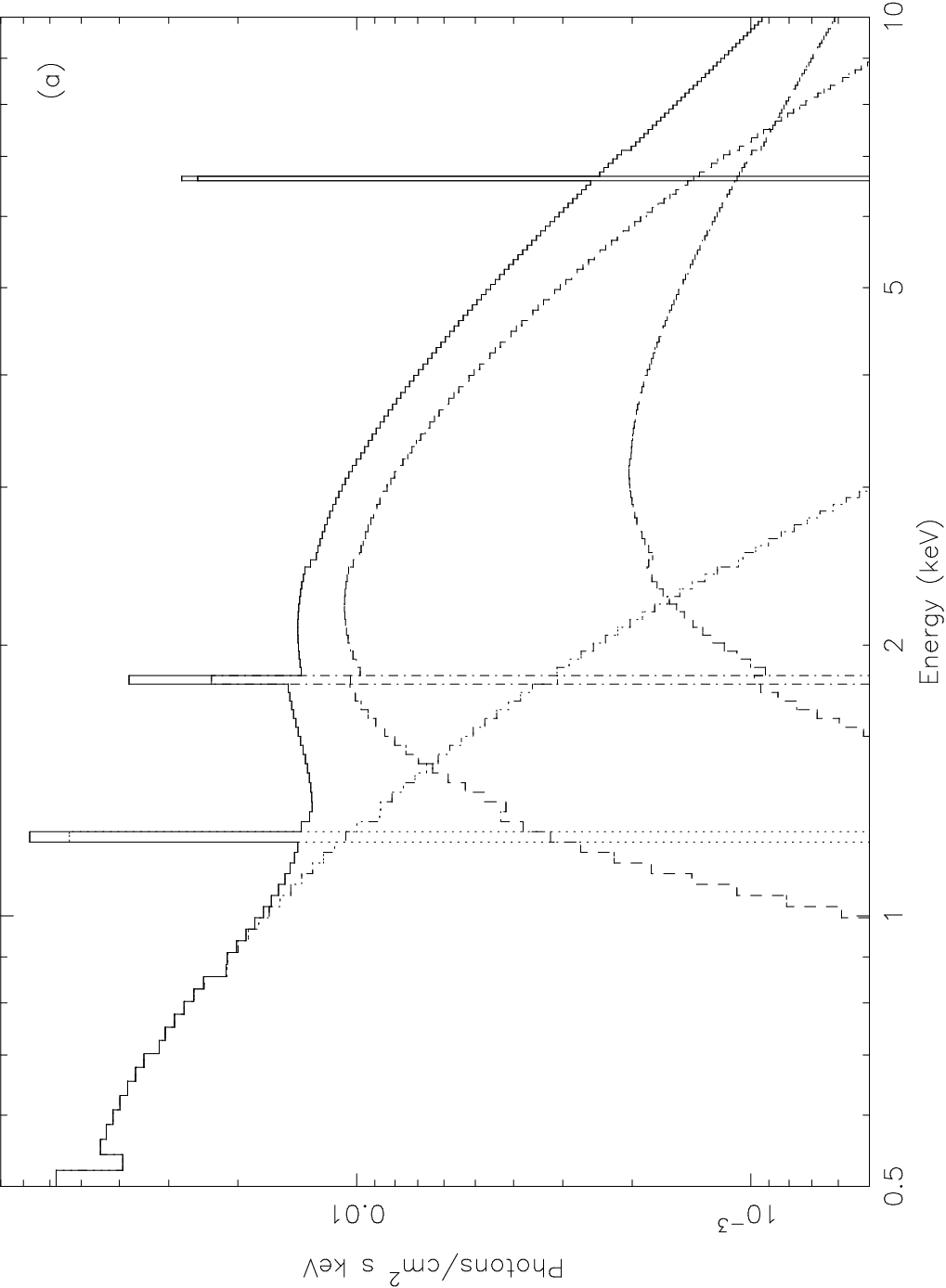}{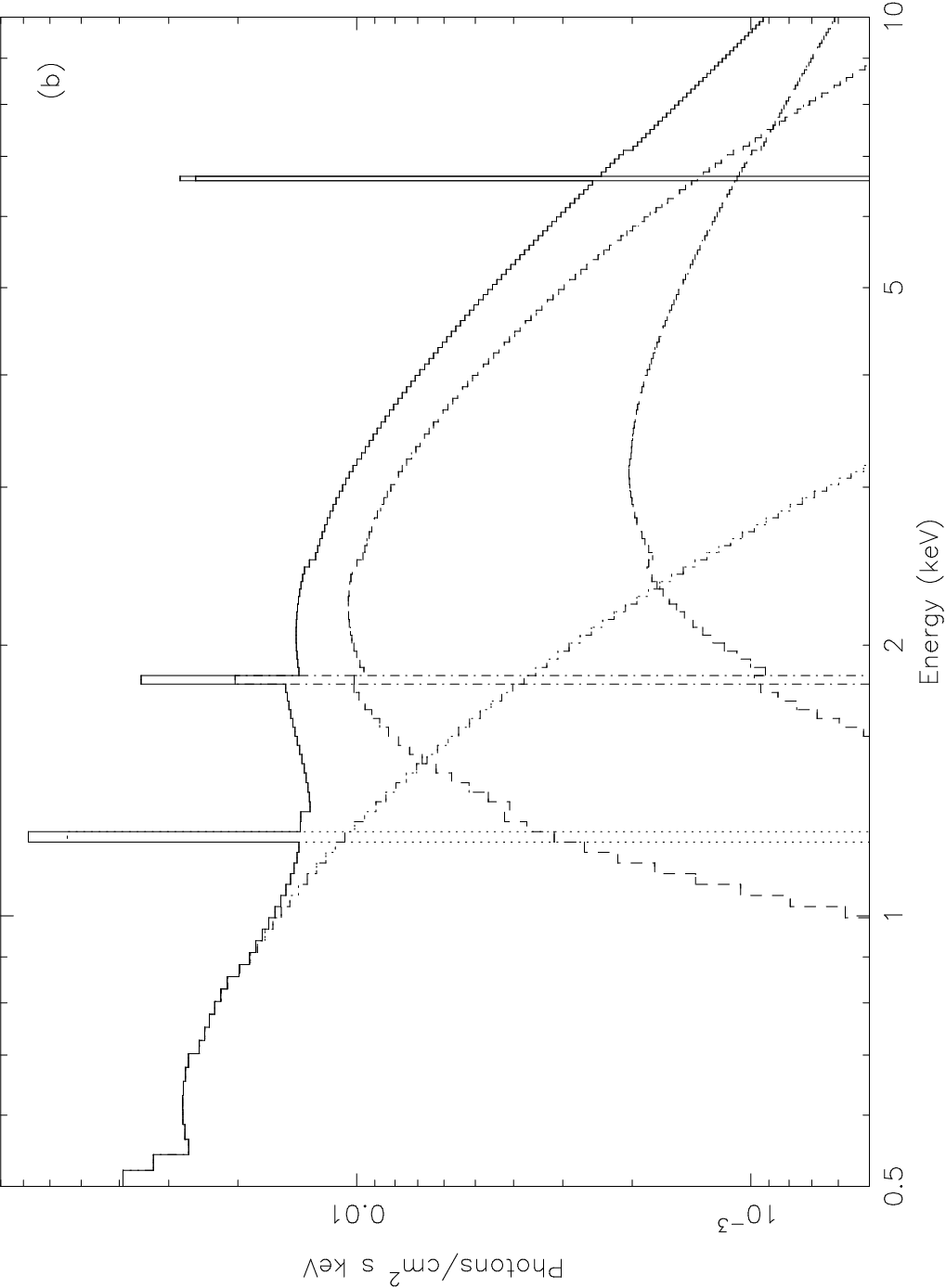}
\figcaption[]{The fitted model spectra with absorbing
column fixed at 1 $\times$ 10$^{21}$ cm$^{-2}$ for the
soft component. All other parameters are fixed at the 
best fit values as presented in Section \S \ref{subsec:line}. 
Panel (a) is with a continuous absorption and (b) is with 
a single absorption for the soft component.
\label{fig:fig7}}
\end{figure}

\begin{figure}[]
\figurenum{8}
\epsscale{1.0}
\plottwo{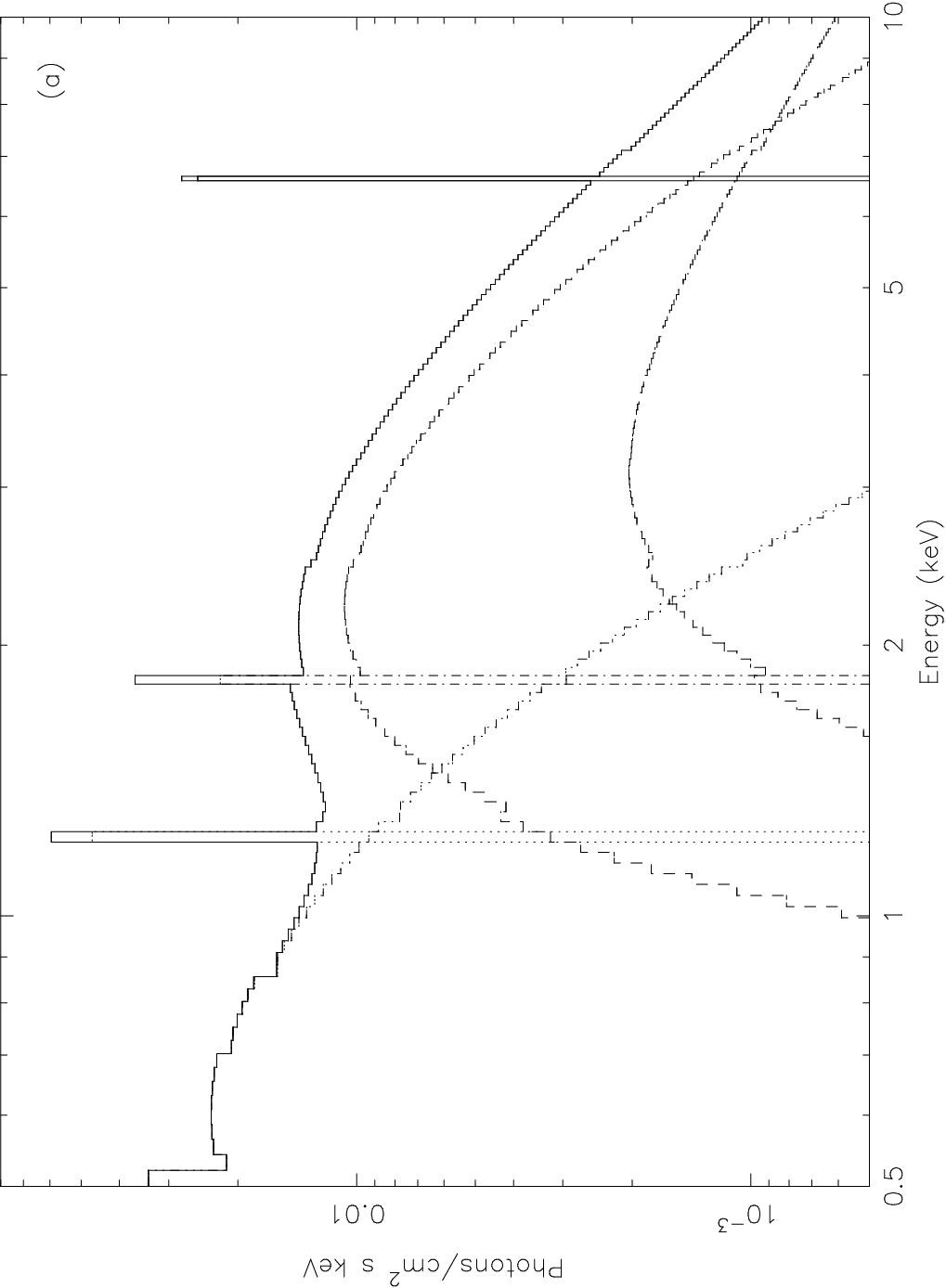}{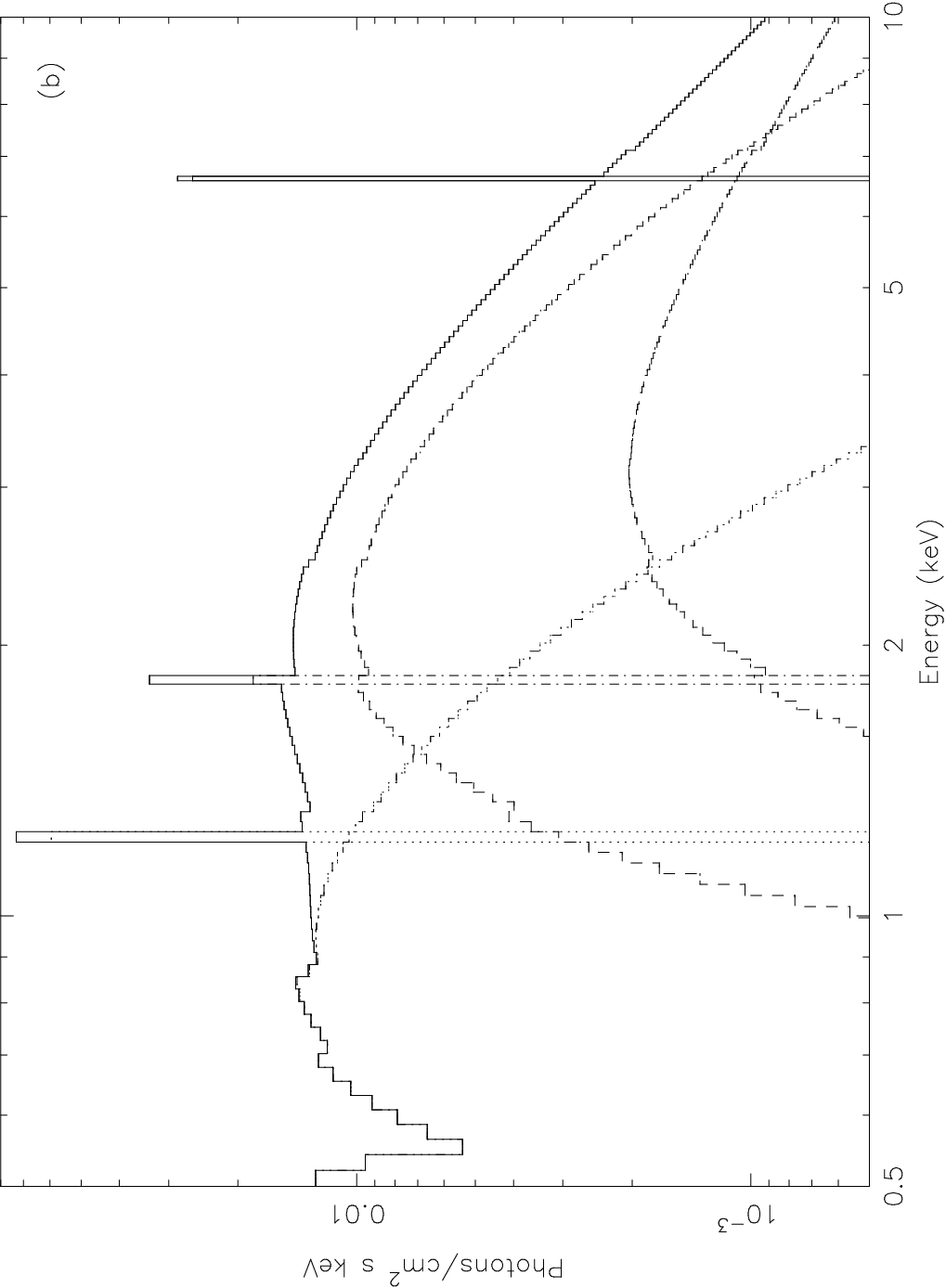}
\figcaption[]{The fitted model spectra with absorbing
column fixed at 3 $\times$ 10$^{21}$ cm$^{-2}$ for the
soft component. All other parameters are fixed at the 
best fit values as presented in Section \S \ref{subsec:line}. 
Panel (a) is with a continuous absorption and (b) is with 
a single absorption for the soft component.
\label{fig:fig8}}
\end{figure}

Although the model fitting with the {\it ASCA} data is not sensitive
to the low columns ($<$5 $\times$ 10$^{21}$ cm$^{-2}$) providing
only an upper limit, we may further constrain the column density
by using {\it ROSAT} results. Figure~\ref{fig:fig6} compares
the modeled 1.5 to 0.75 keV hardness ratios with the {\it ROSAT} data.
Since the absorbing column is not constrained, the modeled hardness ratios
are presented in a range within the upper limit ($<$5 $\times$ 
10$^{21}$ cm$^{-2}$ for both continous and single absorptions). 
The measured hardness ratio of the {\it ROSAT} data
for the same region is $\sim$1.4 (P98). The uncertainty of the {\it ROSAT}
measurement (Note: this measurement is independent of the plotted N$_H$ 
range in Figure~\ref{fig:fig6}) is displayed as the 
area between the dotted horizontal lines in Figure~\ref{fig:fig6}. 
The {\it ROSAT} measurement of the hardness ratio can be obtained in the 
model with N$_H$ $\sim$ 3 $\times$ 10$^{21}$ cm$^{-2}$ for the continous
absorption and N$_H$ $\sim$ 1 $\times$ 10$^{21}$ cm$^{-2}$ for the single 
absorption. Assuming a space density of 1 cm$^{-3}$, these absorbing 
columns imply a distance of $\la$1 kpc through the soft X-ray emitting 
regions. The direct comparison 
between the modeled flux and the measured {\it ROSAT} flux is complicated
since the {\it ROSAT} data have been obtained from a mosaic of several 
pointings reflecting a mixture of different vignettings and responses. 
In Figure~\ref{fig:fig6}, the overall spectra with the single absorption 
are harder and show a larger divergence from the {\it ROSAT} measurement 
than the continuous absorption model. The harder spectra for the single 
absorption is perhaps as expected since the soft photons are to be more
absorbed by the ``single foreground'' absorbing column where the same 
``total'' absorption is considered as is with the continuous absorption. 
In order to demonstrate the effects of the continuous absorption,
comparisons between the spectral models fitted to our data with the
continuous and the single absorption (for the soft component)
are presented in Figure~\ref{fig:fig7} and Figure~\ref{fig:fig8}.
Figure~\ref{fig:fig7}a and Figure~\ref{fig:fig7}b compare two models
where the ``total'' absorption is fixed at 1 $\times$ 10$^{21}$
cm$^{-2}$ for the soft component while Figure~\ref{fig:fig8}a and
Figure~\ref{fig:fig8}b are where the column is 3 $\times$ 10$^{21}$
cm$^{-2}$. All other parameters are the best fit values to fit
the data used in this paper.
It is evident that the effect of the distribution of the
absorbing material is small with a low column and that is more
significant with a larger column where more soft photons
($<$1 keV) are absorbed with the single absorption.
Since the single absorption with 1 $\times$ 10$^{21}$ cm$^{-2}$
and the continuous absorption with 3 $\times$ 10$^{21}$ cm$^{-2}$
have almost identical spectral shape at even below 1 keV, 
further distinctions between them may be infeasible even by using
the {\it ROSAT} results.
The reality is most likely in between these two 
extreme cases and so these two columns ($\sim$1 $\times$ 10$^{21}$ 
cm$^{-2}$ and $\sim$3 $\times$ 10$^{21}$ cm$^{-2}$) may represent a 
range of the actual absorption through the soft X-ray emitting regions.

After deabsorbing the spectrum,
the average values of the modeled 0.5 $-$ 10 keV band surface brightness, 
volume emissivity, and the total luminosity with the $N_H$ in the range 
of 1 $-$ 3 $\times$ 10$^{21}$ cm$^{-2}$ are 0.27 $\times$ 10$^{-7}$ ergs 
s$^{-1}$ cm$^{-2}$ sr$^{-1}$, 1.11 $\times$ 10$^{-28}$ ergs s$^{-1}$ cm$^{-3}$,
and 0.2 $\times$ 10$^{37}$ ergs s$^{-1}$, respectively.
The modeled average surface brightness in the 0.5 $-$ 10 keV band
($\sim$0.27 $\times$ 10$^{-7}$ ergs s$^{-1}$ cm$^{-2}$ 
sr$^{-1}$) of the unabsorbed soft component is 
substantially lower than the previous result at {\it b} = 0$^{\circ}$
toward the Scutum arm ($\sim$2 $\times$ 10$^{-6}$ ergs s$^{-1}$ 
cm$^{-2}$ sr$^{-1}$) \citep{kel97}. It may be due to the 
``high'' latitude of the current data (centered on {\it b} = 
$\sim$0.8$^{\circ}$) as the DXB near the plane has significant
flux variations along the latitudes~\citep{kel97}.
The low surface brightness may also indicate a soft X-ray flux variation 
along the Galactic longitudes. 
With the estimated volume emissivity the mean electron density of the
plasma can be calculated by the relation, $\epsilon$ = $n_e^{2}$$\Lambda$(T),
where $\epsilon$ is the emissivity, and $n_e$ is the electron density.
The cooling function $\Lambda$(T) is estimated to be,
$${\Lambda}(T)~=~6.2~{\times}~10^{-19}~T^{-0.6},~0.009~keV~{<}~
kT~{<}~3.4~keV$$
$${\Lambda}(T)~=~2.5~{\times}~10^{-27}~T^{0.5},~kT~{>}~3.4~keV$$
\citep{mc77}. The estimated electron density is  0.0017 cm$^{-3}$. 
The thermal pressure ($p/k$) can then be estimated as $p/k$ = 2$n_e$T 
assuming the adiabatic phase of the plasma. 
In Table~\ref{tbl:tab2}, the modeled parameters with the $N_H$ in the range 
of 1 (the single absorption) $-$ 3 (the continuous absorption) 
$\times$ 10$^{21}$ cm$^{-2}$ are presented as
an average value of each parameter for the given range of the $N_H$ column.

\subsection{\label{subsec:hard} The Hard Component}

A hard component with {\it kT} = 5 $-$ 10 keV ($\sim$8 keV on average)
has been reported for the GRXE \citep{kel86a,kel97,yel97}. 
The present result implies that the hard component of the Galactic X-ray 
background in the plane is from a {\it kT} = $\sim$3.7 keV plasma.
While the uncertainty associated with this temperature is large,
it is still relatively lower than the previous results.
The difference in the hard component temperature may be due to
the assumptions in the model fittings. We have assumed an ionization
equilibrium for the thermal plasma and no contribution from 
a power law hard tail in the model. 
A broad range of the plasma temperature for the hard component has 
been obtained with various data depending on different assumptions 
within the model fittings (see \S \ref{rsmodel}). 
With the current {\it ASCA} data it is however
infeasible to test different models and assumptions with meaningful
statistics leaving such investigations in the future works.

The origin of the ``lower'' temperature might also be associated with the 
complicated Galactic structure in the {\it l} $\sim$ 25$^{\circ}$ 
direction. The {\it l} $\sim$ 25$^{\circ}$ direction of the plane has 
been suggested to be a ``transition'' region between the GXB and the 
molecular ring (P98). 
The angular extension of the molecular ring ($-$60$^{\circ}$ $<$ {\it l} 
$<$ 60$^{\circ}$) \citep{del87} is coincident with that of 
the GRXE. The molecular ring may trace the active star-forming regions
in the Galaxy and be spatially correlated with the emitting regions of
the GRXE as suggested by Yamauchi \& Koyama (1993).
The estimated temperature of the GXB plasma is most
likely {\it kT} = 0.35 keV \citep{sel97,p98}.
Assuming that the {\it l} $\sim$ 25$^{\circ}$ direction beyond the
absorbing cloud is a transition region between the GXB and the
hard component GRXE, the relatively lower temperature
of the present data may reflect the existence of the ``cooler''
GXB plasma along the line of sight. Observations at various longitudes 
along the plane will be necessary to test this speculation.

Assuming that the hard component occupies a thin disk along the plane
(with a thickness of $\sim$ 200 pc, and a radius of $\sim$ 10 kpc),
the 0.5 $-$ 10 keV total luminosity is $\sim$1.0 $\times$ 10$^{38}$ ergs 
s$^{-1}$. This total luminosity is in good agreement with 
the previous results for the hard component GRXE ($\sim$1 $-$ 2 $\times$
10$^{38}$ ergs s$^{-1}$) \citep{wel82,kel86a,kel97,vm98}.
The estimated electron density is $\sim$0.0018 cm$^{-3}$ and the thermal
pressure is $p/k$ = 1.5 $\times$ 10$^{5}$ K cm$^{-3}$.

\section{\label{sec:summary} SUMMARY AND CONCLUSIONS}

The Galactic X-ray background emission in the direction of {\it l, b}
= 25.6$^{\circ}$, 0.78$^{\circ}$ is observed with \asca. 
The observed pointing direction is toward a dense molecular cloud 
which casts an X-ray shadow in the 0.5 $-$ 2 keV band. The 
kinematic distance to the cloud (3 kpc) and the presence
of the soft X-ray shadow makes it feasible to study the spectral 
properties of the Galactic X-ray background emission originating in the 
foreground of the Galactic X-ray bulge yet beyond the Local Hot Bubble. 

The observed spectrum can be fitted with
a two component thermal plasma model while candidate atomic emission lines 
are separately fitted with narrow Gaussian profiles.
The soft component is from a {\it kT} = $\sim$1 keV plasma with 
low absorption and dominates the observed spectrum at $<$2 keV. 
When compared with the {\it ROSAT} data, the absorbing column for the
soft component is most likely in the range of $\sim$1 $-$ 3 $\times$ 10$^{21}$
cm$^{-2}$ depending on the specific distributions of the absorbing
material along the line of sight.
This range of the absorbing column implies a 
distance scale of $\la$ 1 kpc through the soft component X-ray emitting regions.
These results indicate that the 0.5 $-$ 2 keV soft X-ray background emission 
in the Galactic plane arises from an $\sim$10$^{7}$ K thermal
plasma within $\sim$1 kpc from the Sun, at least in the {\it l} $\sim$
25$^{\circ}$ direction.

The hard component is from a {\it kT} = $\sim$3.7 keV plasma and absorbed
by the Galactic disk. The best fit temperature is relatively lower
compared to the previous studies of the GRXE. The difference in the
continuum temperature may reflect the assumptions in the given model fitting
or may be due to the complexity of the Galactic DXB emitting regions along 
the line of sight. 
Follow up observations of the Galactic DXB for various directions along the
plane with {\it XMM-Newton} and/or {\it Chandra} will be necessary to
address the complicated nature of the Galactic DXB along the plane.

\acknowledgments

{The authors thank S. Snowden for valuable discussion, and
A. Valinia for helpful comments at the early stage of this work.
We also thank the referee for the constructive comments.
This work was supported by USRA and NASA Cooperative Agreement 
NCC 5-356.}

\clearpage

\begin{deluxetable}{cccccc}
\footnotesize
\tablecaption{Best fit parameters for bremsstrahlung + Gaussian model.
\label{tbl:tab1}}
\tablewidth{0pt}
\tablehead{ & \colhead{Continuous} & \colhead{absorption} && \colhead{Single}
 & \colhead{absorption}\\
\hline
 & \colhead{$N_{H}$} & \colhead{$kT$}&&
\colhead{$N_{H}$} & \colhead{$kT$}\\
 & \colhead{(10$^{22}$ cm$^{-2}$)} & \colhead{(keV)} &&
\colhead{(10$^{22}$ cm$^{-2}$)} & \colhead{(keV)}}
\startdata
Soft & $<$0.48\tablenotemark{a} & 0.92$^{+0.41}_{-0.39}$ &
& $<$0.51\tablenotemark{a} &1.02$^{+0.47}_{-0.46}$ \\
Hard & 2.1\tablenotemark{b} & 3.67\tablenotemark{b}& &
2.1\tablenotemark{b} & 3.67\tablenotemark{b} \\
$\chi$$^{2}_{\nu}$=0.45 & & &$\chi$$^{2}_{\nu}$=0.45 & &\\
\enddata

\tablenotetext{a}{This is a 2$\sigma$ upper limit.} 

\tablenotetext{b}{This is the best fit value from the hard component
fit in the 3 $-$ 9 keV band and is fixed in the overall fitting.  }

\end{deluxetable}

\clearpage

\begin{deluxetable}{lcc}
\footnotesize
\tablecaption{Summary of modeled plasma parameters for the
soft and hard components after removing the absorption.
\label{tbl:tab2}}
\tablewidth{0pt}
\tablehead{
\colhead{Parameter} & \colhead{Soft\tablenotemark{a,b}} &
\colhead{Hard\tablenotemark{a}} }
\startdata
Surface Brightness (ergs s$^{-1}$ cm$^{-2}$ sr$^{-1}$) & 0.27 $\times$
10$^{-7}$ & 1.89 $\times$ 10$^{-7}$  \\
Volume Emissivity (ergs s$^{-1}$ cm$^{-3}$) & 1.11 $\times$
10$^{-28}$ & 0.51 $\times$ 10$^{-28}$ \\
Total Luminosity (ergs s$^{-1}$) & 0.2 $\times$ 10$^{37}$ &
0.94 $\times$ 10$^{38}$ \\
Electron Density (cm$^{-3}$) & 0.0017 & 0.0018 \\
Thermal Pressure ({\it p/k}, K cm$^{-3}$) & 3.6 $\times$ 10$^{4}$ &
1.5 $\times$ 10$^{5}$ \\
\enddata

\tablenotetext{a}{These are derived from the 0.5 $-$ 10.0 keV band.}

\tablenotetext{b}{These are the average values for each model parameter
                  where $N_H$ is in the range of 1 $-$ 3 $\times$ 10$^{21}$
                  cm$^{-2}$.}


\end{deluxetable}

\clearpage

\begin{thebibliography}{}

\bibitem[Almy et al. 2000]{ael00} Almy, R. C., McCammon, D., Digel, S. W.,
Bronfman, L., \& May, J. 2000, ApJ, in press
\bibitem[Cox \& Reynolds 1987]{cr87} Cox, D. P. \& Reynolds, R. J. 1987,
ARA\&A, 25, 303
\bibitem[Chen et al. 1997]{cfg97} Chen, L.-W, Fabian, A.C., \&
Gendreau, K. C. 1997, MNRAS, 285, 449
\bibitem[Dame et al. 1987]{del87} Dame, T. M., Ungerechts, H., Cohen,
     R.S., de Geus, E. J., Grenier, I. A., May, J., Murphy, D. C.,
      Nyman, L. -\AA, \& Thaddeus, P. 1987, ApJ, 322, 706
\bibitem[Gendreau 1995]{gen95} Gendreau, K. C. 1995, Ph.D. Thesis, 
Massachusetts Institute of Technology
\bibitem[Ishisaki 1996]{ish96} Ishisaki, Y. 1996, Ph.D. Thesis,
University of Tokyo
\bibitem[Jakobsen \& Kahn 1986]{jk86} Jakobsen, P. \& Kahn, S. M. 1986,
ApJ, 309, 682
\bibitem[Kaneda 1996]{k96} Kaneda, H. 1996, Ph.D. Thesis, University of Tokyo
\bibitem[Kaneda et al. 1997]{kel97} Kaneda, H., Makishima, K., Yamauchi, S.
Koyama, K., Matsuzaki, K, \& Yamasaki, N. Y. 1997, ApJ, 491, 638
\bibitem[Koyama et al. 1986a]{kel86a} Koyama, K., Makishima, K., Tanaka, Y.,
\& Tsunemi, H. 1986a, PASJ, 38, 121
\bibitem[Koyama et al. 1986b]{kel86b} Koyama, K., Ikeuchi, S., \&
Tomisaka, K. 1986b, PASJ, 38, 503
\bibitem[McKee \& Cowie 1977]{mc77} McKee, C. F., \& Cowie, L. L. 1977, ApJ,
215, 213
\bibitem[Morrison \& McCammon 1983]{mm83} Morrison, R. \& McCammon, D. 1983,
ApJ, 270, 119
\bibitem[Nousek et al. 1982]{nel82} Nousek, J. A., Fried, P. M., Sanders,
W. T., \& Kraushaar, W. L. 1982, ApJ, 258, 83
\bibitem[Ottmann \& Schmitt 1992]{os92} Ottmann, R., \& Schmitt, J. H. M. M. 
     1992, A\&A, 256, 421
\bibitem[Park et al. 1997]{pel97} Park, S., Finley, J. P., Snowden, S.
L., \& Dame, T. M. 1997, ApJ, 476L, L77 
\bibitem[Park 1998]{p98} Park, S. 1998, Ph.D. Thesis, Purdue University
\bibitem[Park et al. 1998]{pel98} Park, S., Finley, J. P., \&
Dame, T. M. 1998, ApJ, 509, 203 (P98)
\bibitem[Schmitt \& Snowden 1990]{ss90} Schmitt, J. H. M. M. \& 
    Snowden, S. L. 1990, ApJ, 361, 207
\bibitem[Serlemitsos et al. 1995]{sel95} Serlemitsos, P. J.,
Jalota, L., Soong, Y., Kunieda, H., Tawara, Y., Tsusaka, Y.,
Suzuki, H., Sakima, Y., Yamazaki, T., Yoshioka, H., Furuzawa, A.,
Yamashita, K., Awaki, H., Itoh, M., Ogasaka, Y., Honda, H., \&
Uchibori, Y. 1995, PASJ, 47, 105
\bibitem[Smith 1999]{s99} Smith, R., 1999, private communications
\bibitem[Snowden et al. 1994]{sel94} Snowden, S. L., McCammon, D.,
     Burrows, D. N., \& Mendenhall, J. A. 1994, ApJ, 424, 714
\bibitem[Snowden et al. 1997]{sel97} Snowden, S. L., Egger, R., Freyberg, 
     M. J., McCammon, D., Plucinsky, P. P., Sanders, W. T., Schmitt, J. H. 
     M. M., Tr\"umper, J., \& Voges, W., 1997, ApJ, 485, 125
\bibitem[Snowden et al. 1998]{sel98} Snowden, S. L., Egger, R., Finkbeiner, 
D. P., Freyberg, M. J., \& Plucinsky, P. P. 1998, ApJ, 493, 715
\bibitem[Valinia \& Marshall 1998]{vm98} Valinia, A., \& Marshall, F. E. 1998,
ApJ, 505, 134
\bibitem[Valinia et al. 2000]{vel00} Valinia, A., Tatischeff, V., Arnaud, K.,
Ebisawa, K., \& Ramaty, R. 2000, ApJ, in press
\bibitem[Wang 1992]{w92} Wang, Q. D. 1992, ApJ, 392, 509
\bibitem[Warwick et al. 1985]{wel85} Warwick, R. S., Turner, M. J. L., Watson,
M. G., \& Willingale, R. 1985, Nature, 317, 218
\bibitem[Worrall et al. 1982]{wel82} Worrall, D. M., Marshall, F. E.,
Boldt, E. A., \& Swank, J. H. 1982, ApJ, 255, 111
\bibitem[Yamasaki et al. 1997]{ysel97} Yamasaki, N. Y., Ohashi, T.,
Takahara, F., Yamauchi, S., Koyama, K., Kamae, T., Kaneda, H., 
Makishima, K., Sekimoto, Y., Hirayama, M., Takahashi, T., Yamagami, T.,
Gunji, S., Tamura, T., Miyazaki, S., \& Nomachi, M. 1997, ApJ, 481, 821
\bibitem[Yamauchi \& Koyama 1993]{yk93} Yamauchi, S., \& Koyama, K. 1993,
ApJ, 404, 620
\bibitem[Yamauchi \& Koyama 1995]{yk95} Yamauchi, S., \& Koyama, K. 1995,
PASJ, 47, 439
\bibitem[Yamauchi et al. 1997]{yel97} Yamauchi, S., Kaneda, H., Koyama, K.,
Makishima, K., Matsuzaki, K, Sonobe, T., Tanaka, Y., \& Yamasaki, N. Y.
1997, {\it X-ray Imaging and Spectroscopy of Cosmic Hot Plasmas} Proceedings
of International Symposium on X-ray Astronomy, edited by F. Makino \& K.
Mitsuda, (Universal Academy Press, Inc. Tokyo, Japan), 145

\end{thebibliography}
\end{document}